\pdfoutput=1
\documentclass{eptcs}
\providecommand{\event}{Workshop on Quantitative Formal Methods: Theory and Applications} 
\providecommand{\volume}{??}
\providecommand{\anno}{2009}
\providecommand{\firstpage}{1}
\providecommand{\eid}{??}
\usepackage{graphicx,amsmath,verbatim,multirow}

\newcommand{\Clock}{{\sf Clock}}
\newcommand{\Receiver}{{\sf Receiver}}
\newcommand{\Sender}{{\sf Sender}}
\newcommand{\Synchronizer}{{\sf Synchronizer}}
\newcommand{\Controller}{{\sf Controller}}
\newcommand{\id}{{\sf id}}

\newcommand{\Max}{{\sf max}}
\newcommand{\Min}{{\sf min}}
\newcommand{\N}{{\sf N}}
\newcommand{\n}{{\sf n}}
\newcommand{\C}{{\sf C}}
\newcommand{\Nodes}{{\sf Nodes }}
\newcommand{\Guard}{{\sf g}}
\newcommand{\Radioswitchtime}{{\sf r}}

\newcommand{\K}{{\sf k_0}}
\newcommand{\tsn}{{\sf tsn}}

\newcommand{\clk}{{\sf clk}}
\newcommand{\csn}{{\sf csn}}

\newcommand{\Uppaal}{\textsc{Uppaal}}

\title{Modelling Clock Synchronization in the Chess gMAC WSN Protocol%
\thanks{Research supported by the European Community's Seventh Framework Programme under grant agreement no 214755 (QUASIMODO)
and by the DFG/NWO bilateral cooperation project ROCKS.}}
\author{Mathijs Schuts \and Feng Zhu \and Faranak Heidarian%
\thanks{Research supported by NWO/EW project 612.064.610 Abstraction Refinement for Timed Systems (ARTS).}%
\and Frits Vaandrager
\institute{Institute for Computing and Information Sciences\\ Radboud University Nijmegen\\ P.O.\ Box 9010, 6500 GL Nijmegen, The Netherlands}
\email{M.Schuts@student.ru.nl, FengZhu@student.ru.nl, faranak@cs.ru.nl, F.Vaandrager@cs.ru.nl}
}

\def\titlerunning{Modelling Clock Synchronization in the Chess gMAC WSN Protocol}
\def\authorrunning{Schuts, Zhu, Heidarian and Vaandrager}
\begin{document}
\maketitle

\begin{abstract}
We present a detailled timed automata model of the clock synchronization algorithm that is currently being used
in a wireless sensor network (WSN) that has been developed by the Dutch company Chess.  Using the \Uppaal\  model checker,
we establish that in certain cases a static, fully synchronized network may eventually become unsynchronized if
the current algorithm is used, even in a setting with infinitesimal clock drifts.
\end{abstract}
\section{Introduction}
\begin{figure}[h!]
\begin{center}
\includegraphics[scale=0.5]{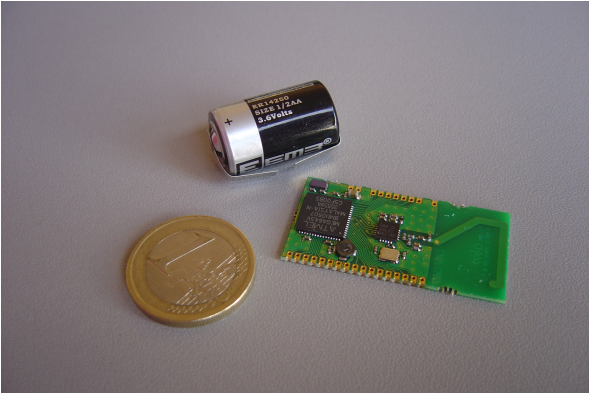}
\end{center}
\caption{Chess MyriaNode 2.4 Ghz wireless sensor node}
\label{Chessnode}
\end{figure}

\begin{figure}[h!]
\begin{center}
\includegraphics[scale=0.25]{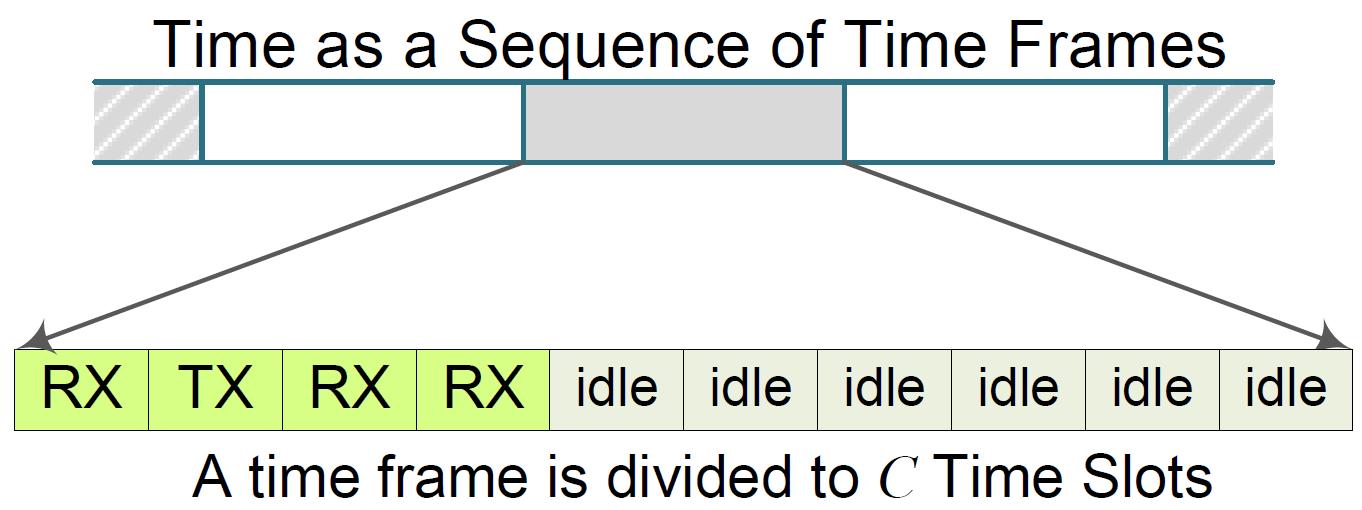}
\end{center}
\caption{The structure of a time frame}
\label{fig:TimeModel}
\end{figure}

Wireless sensor networks consist of autonomous devices that communicate via radio and use sensors to cooperatively monitor
physical or environmental conditions.
In this paper, we formally model and analyze a distributed algorithm for clock synchronization in wireless sensor networks
that has been developed by the Dutch company Chess in the context of the MyriaNed project \cite{del52}.
Figure~\ref{Chessnode} displays a sensor node developed by Chess.
The algorithm that we consider is part of the \emph{Medium Access Control (MAC) layer}, which is responsible for the access to the
wireless shared channel.
Within its so-called gMAC protocol, Chess uses a Time Division Multiple Access (TDMA) protocol.
Time is divided in fixed length \emph{frames}, and each frame is subdivided
into \emph{slots} (see Figure~\ref{fig:TimeModel}).
Slots can be either \emph{active} or \emph{sleeping} (\emph{idle}). During active slots, a node is either listening for incoming messages from
neighboring nodes (\emph{RX}) or it is sending a message (\emph{TX}).  During sleeping slots a node is switched to energy saving mode.
Since energy efficiency is a major concern in the design of wireless sensor networks,
the number of active slots is typically much smaller than the total number of slots
(less than $1\%$ in the current implementation).
The active slots are placed in one contiguous sequence which currently is placed at the beginning of the frame.
A node can only transmit a single message per time frame, during its TX slot. The protocol takes care that neighboring nodes have different TX slots.

One of the greatest challenges in the design of the MAC layer is to find suitable mechanisms for clock synchronization:
we must ensure that whenever some node is sending all its neighbors are listening.
Sensor nodes come equipped with a crystal clock, which may  drift. This may cause the TDMA time slot boundaries to drift and
thus lead to situations in which nodes get out of sync. To overcome this problem nodes will have to adjust their clocks now and then.
Also, the notion of \emph{guard time} is introduced: at the beginning of its TX slot, a sender waits a certain amount of time to ensure
that all its neighbors are ready to receive messages.  Similarly, a sender does not transmit for a certain amount of time at the end of its TX slot.
In order to save energy it is important to reduce these guard times to a minimum.
Many clock synchronization protocols have been proposed for wireless sensor networks,
see e.g.\  \cite{Sun05,FL06,Tjoa04,MT05,Assegei08,LenzenLW08,Pussente09}.
However, these protocols (with the exception of \cite{Tjoa04,Assegei08} and possibly \cite{Pussente09})
involve a computation and/or communication overhead that is
unacceptable given the extremely limited resources (energy, memory, clock cycles) available within the Chess nodes.

To experiment with its designs, Chess currently builds prototypes and uses advanced simulation tools.  However, due to the huge number
of possible network topologies and clock speeds of nodes, it is difficult to discover flaws in the clock synchronization algorithm via
these methods.  
Timed automata model checking has been succesfully used for the analysis of worst case scenarios for protocols that involve
clock synchronization, see for instance \cite{BenEtAl96,HavelundSLL97,VG06}.  To enable model checking, models need to be much more
abstract than for simulation, and also the size of networks that can be tackled is much smaller, but the big advantage is that
the full state space of the model can be explored.

In this paper, we present a detailed model of the Chess gMAC algorithm using the input language of the timed automata model checking
tool \Uppaal\  \cite{BehrmannDL04}.
Another \Uppaal\  model for the gMAC algorithm is presented in \cite{HSV09}, but that model deviates and abstracts from several
aspects in the implementation in order to make verification feasible.  The aim of the present paper is to construct a model that
comes as close as possible to the specification of the clock synchronization algorithm presented in \cite{del52}. 
Nevertheless, our model still does not incorporate some features of the full algorithm and network, such as dynamic slot allocation,
synchronization messages, uncertain communication delays, and unreliable radio communication.
At places where the informal specification of \cite{del52} was incomplete or ambiguous, the engineers from Chess kindly provided us with
additional information on the way these issues are resolved in the current implementation of the network \cite{communication}.
In the current implementation of Chess, a node can only adjust its clock once every time frame during the sleeping period, using an
extension of the Median algorithm of \cite{Tjoa04}.  This contrasts with the approach in \cite{HSV09} in which a sensor node may adjust its clock after every received message.
In the present paper we faithfully model the Median algorithm as implemented by Chess.
Another feature of the gMAC algorithm that was not addressed in \cite{HSV09} but that we model in this paper is the radio switching time:
there is some time involved in the transition from sending mode to receiving mode (and vice versa), which in some cases may affect the
correctness of the algorithm.

The Median algorithm works reasonably well in practice, but by means of simulation experiments, Assegei \cite{Assegei08} already exposed some
flaws in the algorithm: in some test cases where new nodes join or networks merge, the algorithm fails to converge or nodes may stay out of
sync for a certain period of time.  Our analysis with \Uppaal\  confirms these results.
In fact, we show that the situation is even worse: in certain cases a static, fully synchronized network may eventually become unsynchronized
if the Median algorithm is used, even in a setting with infinitesimal clock drifts.

In Section~\ref{protocol}, we explain the gMAC algorithm in more detail.  Section~\ref{sec:model} describes our \Uppaal\  model of gMAC. 
In Section~\ref{analysis-results}, the analysis results are described.  Finally, in Section~\ref{conclusions}, we draw some conclusions.
In this paper, we assume that the reader has some basic knowledge of the timed automaton tool \Uppaal.
For a detailed account of \Uppaal, we refer to \cite{BehrmannDL04,Uppaal4.0} and to \url{http://www.uppaal.com}.
The \Uppaal\ model described in this paper is available at \url{http://www.mbsd.cs.ru.nl/publications/papers/fvaan/chess09/}.

\section{The gMAC Protocol}
\label{protocol}

In this section we provide additional details about the gMAC protocol as it has currently been implemented by Chess.

\subsection{The Synchronization Algorithm}\label{sec:synchronization}
In each frame, each node broadcasts one message to its neighbors.  The timing of this message is used for synchronization purposes:
a receiver may estimate the clock value of a sender based on the time when the message is received.
Thus there is no need to send around (logical) clock values.
In the current implementation of Chess, clock synchronization is performed once per frame using the following algorithm \cite{Assegei08,communication}:

\begin{enumerate}
	\item In its sending slot, a node broadcasts a packet which contains its transmission slot number.
	\item Whenever a node receives a message it computes the {\tt phase error}, that is the difference (number of clock cycles) between the expected receiving time and the actual receiving time of the incoming message. Note that the difference between the sender's sending slot number (which is also the current slot number of the sender) and the current slot number of the receiving node must also be taken into account when calculating the phase errors.
	\item After the last active slot of each frame, a node calculates the {\tt offset} from the phase errors of all incoming messages in this frame with the following algorithm: 
\begin{verbatim}
     if (number of received messages == 0)
              offset = 0;
     else if (number of received messages <= 2)
              offset = the phase error of the first received message * gain;
     else
              offset = the median of all phase errors * gain
\end{verbatim}
Here {\tt gain} is a coefficient with value 0.5, used to prevent oscillation of the clock adjustment.
\item During the sleeping period, the local clock of each node is adjusted by the computed {\tt offset} obtained from step 3.
\end{enumerate}

In situations when two networks join, it is possible that the phases of these networks differ so much that the nodes in one network
are in active slots whereas the nodes in the other network are in sleeping slots and vice versa. In this case, no messages can be
exchanged between two networks. Therefore in the Chess design, a node will send an extra message in one (randomly selected) sleeping slot to
increase the chance that networks can communicate and synchronize with each other. This slot is called the synchronization slot and
the message is in the same format as in the transmission slot. 
The extreme value of {\tt offset} can be obtained when two networks join: it may occur that the {\tt offset} is larger than half the total number of clock
cycles of sleeping slots in a frame. Chess uses another algorithm called {\tt join} to handle this extreme case. We decided not to model joining of
networks and synchronization messages 
because currently we do not have enough information about the {\tt join} algorithm.

\subsection{Guard Time}\label{sec:guardtime}
\begin{figure}[hpt]
	\begin{center}
		\includegraphics[width = 14cm]{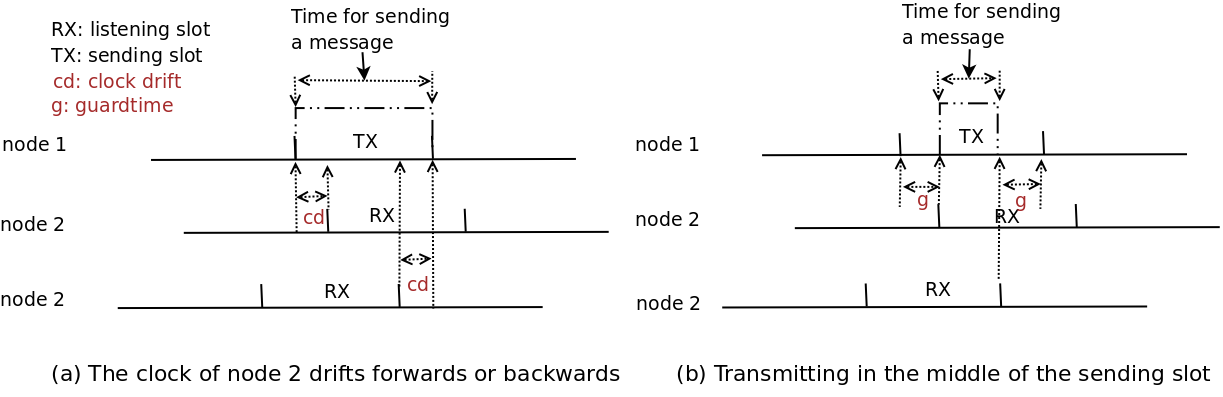}
		\caption{The need for introducing guard times}
		\label{fig:guardtime}
	\end{center}
\end{figure}

The correctness condition for gMAC that we would like to establish is that whenever a node is sending all its neighbors are in receiving mode.
However, at the moment when a node enters its TX slot we cannot guarantee, due to the phase errors, that its neighbors have entered the corresponding
RX slot.
This problem is illustrated in Figure~\ref{fig:guardtime} (a).
Given two nodes 1 and 2, if a message is transmitted during the entire sending slot of node 1
then this message may not be successfully received by node 2 because of the imperfect slot alignment.
Taking the clock of node 1 as a reference, the clock of node 2 may drift backwards or forwards.
In this situation, node 1 and node 2 may have a different view of the current slot number within the time interval where node 1 is sending a message.

To cope with this problem, messages are not transmitted during the entire sending slot but only in the middle, as illustrated in Figure~\ref{fig:guardtime} (b). Both at the beginning and at the end of its sending slot, node 1 does not transmit for a preset period of $\Guard$ clock ticks, in order to accomodate the forwards and backwards clock drift of node 2. Therefore, the time available for transmission equals the total length of the slot minus $2 \Guard$ clock ticks.


\subsection{Radio Switching Time}
\label{sec:radio-switching-time}
The radio of a wireless sensor node can either be in sending mode, or in receiving mode, or in idle mode.
Switching from one mode to another takes time.  In the current implementation of the Chess gMAC protocol, the radio switching time is around
$130 \mu$sec.  The time between clock ticks is around $30 \mu$sec and the guard time $\Guard$ is $9$ clock ticks.
Hence, in the current implementation the radio switching time is smaller than the guard time, but this may change in future implementations.
If the first slot in a frame is an RX slot, then the radio is switched to receiving mode some time before the start of the frame to ensure
that the radio will receive during the full first slot.
However if there is an RX slot after the TX slot then, in order to keep the implementation simple, the radio is switched to the
receiving mode only at the start of the RX slot.
Therefore messages arriving in such receiving slots may not be fully received.
This issue may also affect the performance of the synchronization algorithm.

\section{Uppaal Model}
\label{sec:model}

In this section, we describe the \Uppaal\  model that we constructed of the gMAC protocol.

\begin{figure}[hpt]
		\begin{center}
		\includegraphics[width = 9cm]{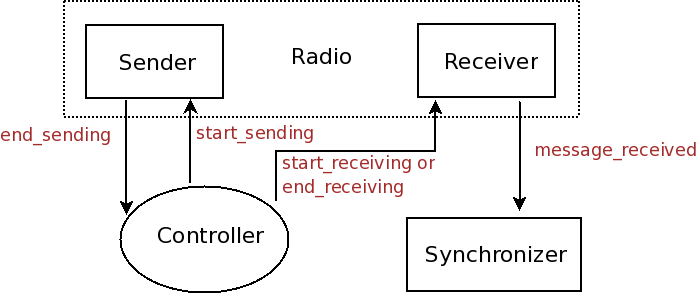}
		\caption{Message flow in the model}
		\label{fig:model}
		\end{center}
\end{figure}

We assume a finite, fixed set of wireless nodes $\Nodes = \{ 0 ,\ldots, \N -1 \}$.
The behavior of an individual node $\id \in \Nodes$ is described by five timed automata
$\Clock(\id)$, $\Receiver(\id)$, $\Sender(\id)$, $\Synchronizer(\id)$ and $\Controller(\id)$.
Figure~\ref{fig:model} shows how these automata are interrelated.  All components
interact with the clock, although this is not shown in Figure~\ref{fig:model}.
Automaton $\Clock(\id)$ models the hardware clock of node $\id$,
automaton $\Sender(\id)$ the sending of messages by the radio,
automaton $\Receiver(\id)$ the receiving part of the radio,
automaton $\Synchronizer(\id)$ the synchronization of the hardware clock, and
automaton $\Controller(\id)$ the control of the radio and the clock synchronization.

Table \ref{tab:NPC} lists the parameters that are used in the model (constants in \Uppaal\ terminology), together
with some basic constraints. The domain of all parameters is the set of natural numbers.
We will now describe the five automaton templates used in our model.
\begin{table}[h!]
        \centering
        \begin{tabular}{l|l|l}
        \hline \hline
        Parameter & Description & Constraints \\
        \hline
        $\N$ & number of nodes & $0<\N$ \\
        $\C$ & number of slots in a time frame & $0<\C$ \\
        $\n$ & number of active slots in a time frame & $0< \n\leq \C$ \\
        $\tsn[\id]$ & TX slot number for node $\id$ & $0\leq \tsn[\id]<\n$\\
        $\K$ & number of clock ticks in a time slot & $0< \K$ \\
        $\Guard$ & guard time & $0< \Guard$ \\
        $\Radioswitchtime$ & radio switch time & $0 \leq \Radioswitchtime$\\
        $\Min[\id]$ & minimal time  between two clock ticks of node $\id$ & $0<\Min[\id]$ \\
        $\Max[\id]$ & maximal time  between two clock ticks of node $\id$ & $\Min[\id] \leq \Max[\id]$ \\
        \hline \hline
        \end{tabular}
        \caption{Protocol parameters}
        \label{tab:NPC}
\end{table}

\begin{figure}[hpt]
		\begin{center}
		\includegraphics[width = 10cm]{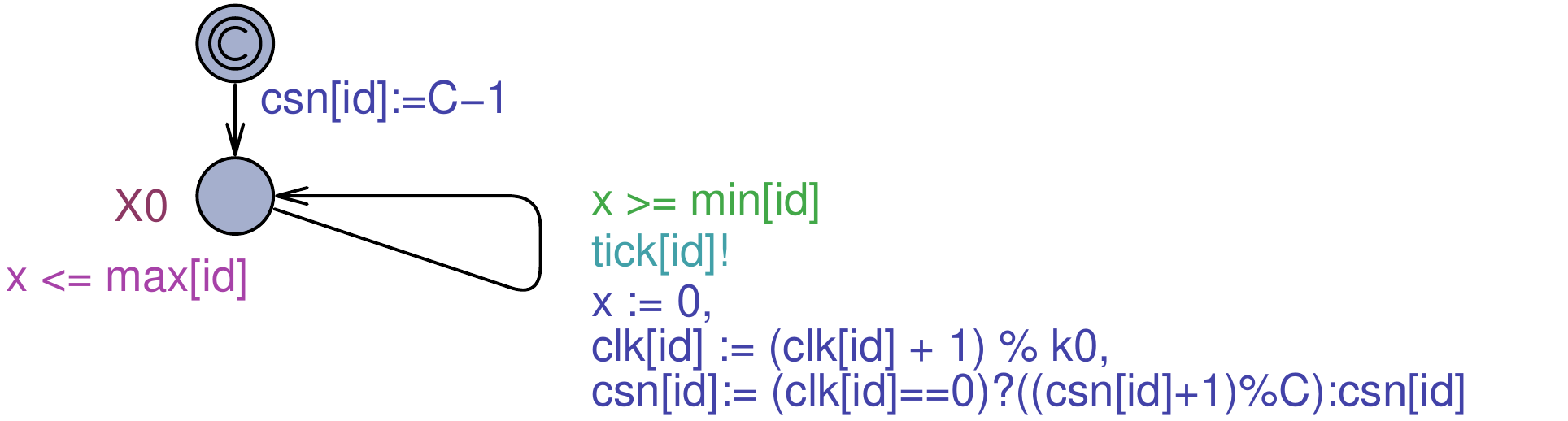}
		\caption{Automaton $\Clock[\id]$}
		\label{fig:wsn-clock}
		\end{center}
\end{figure}

\paragraph{Clock}
Timed automaton $\Clock(\id)$ models the behavior of the hardware clock of node $\id$. The automaton is shown in Figure~\ref{fig:wsn-clock}.
At the start of the system state variable $\csn[\id]$, that records the current slot number,
is initialized to $\C -1$, that is, to the last sleeping slot.
Hardware clocks are not perfect and so we assume a minimal time $\min[\id]$ and a maximal time $\max[\id]$ between successive clock ticks. Integer
variable $\clk[\id]$ records the current value of the hardware clock.  For convenience (and to reduce the size of the state space),
we assume that the hardware clock is reset at the end of each slot, that is after $\K$ clock ticks.
Also, a state variable $\csn[\id]$, which records the current slot number of node $\id$, is updated each time at the start of a new slot.

\begin{figure}[ht!]
		\begin{center}
			\includegraphics[width = 11cm]{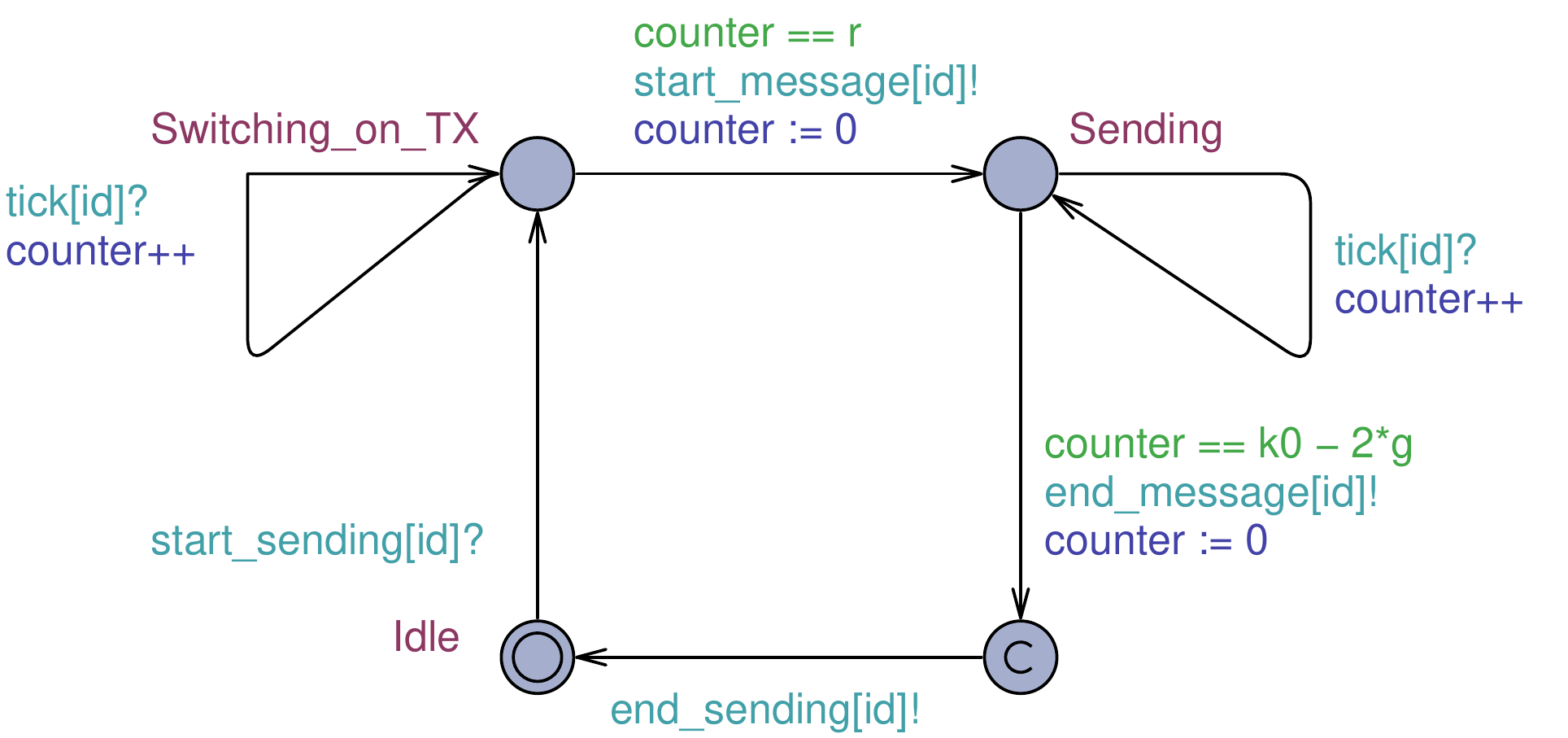}
			\caption{Automaton $\Sender[\id]$}
			\label{fig:wsn-sender}
		\end{center}
\end{figure}

\paragraph{Sender} The sending behavior of the radio is described by the automaton $\Sender[\id]$ shown in Figure \ref{fig:wsn-sender}.
The behavior is rather simple. When the controller asks the sender to transmit a message (via a $\sf start\_sending[\id]$ signal),
the radio first switches to sending mode (this takes $\Radioswitchtime$ clock ticks)
and then transmits the message (this takes $\K - 2 \cdot \Guard$ ticks).
Immediately after the message transmission has been completed, an $\sf end\_sending[\id]$ signal is sent to the controller
to indicate that the message has been sent.

\begin{figure}[hpt]
		\begin{center}
			\includegraphics[width = 11cm]{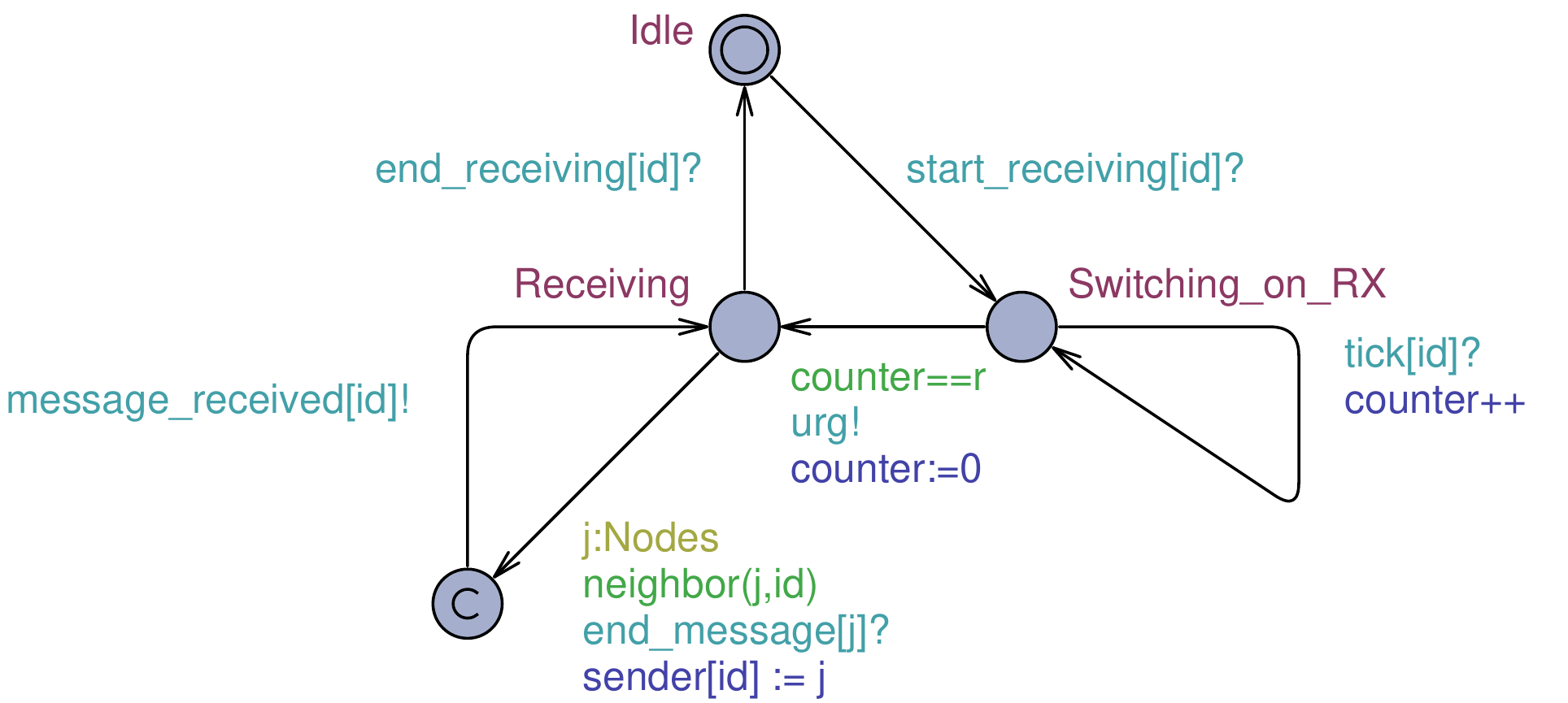}
			\caption{Automaton $\Receiver[\id]$}
			\label{fig:wsn-receiver}
		\end{center}
\end{figure}

\paragraph{Receiver} The automaton $\Receiver[\id]$ models the receiving behavior of the radio.
The automaton is shown in Figure \ref{fig:wsn-receiver}.
Again the behavior is rather simple.  When the controller asks the receiver to start receiving, the receiver first switches to receiving mode
(this takes $\Radioswitchtime$ ticks).  After that, the receiver may receive messages from all its neighbors.  A function $\sf neighbor$ is used
to encode the topology of the network: $\sf neighbor(j,id)$ holds if messages sent by $\sf j$ can be received by $\id$.
Whenever the receiver detects the end of a message transmission by one of its neighbors,
it immediately informs the synchronizer via a $\sf message\_received[\id]$ signal.
At any moment, the controller can switch off the receiver via an $\sf end\_receiving[\id]$ signal.

\begin{figure}[hpt]
	\begin{minipage}[b]{0.5\linewidth}
	\begin{center}
		\includegraphics[width = 8cm]{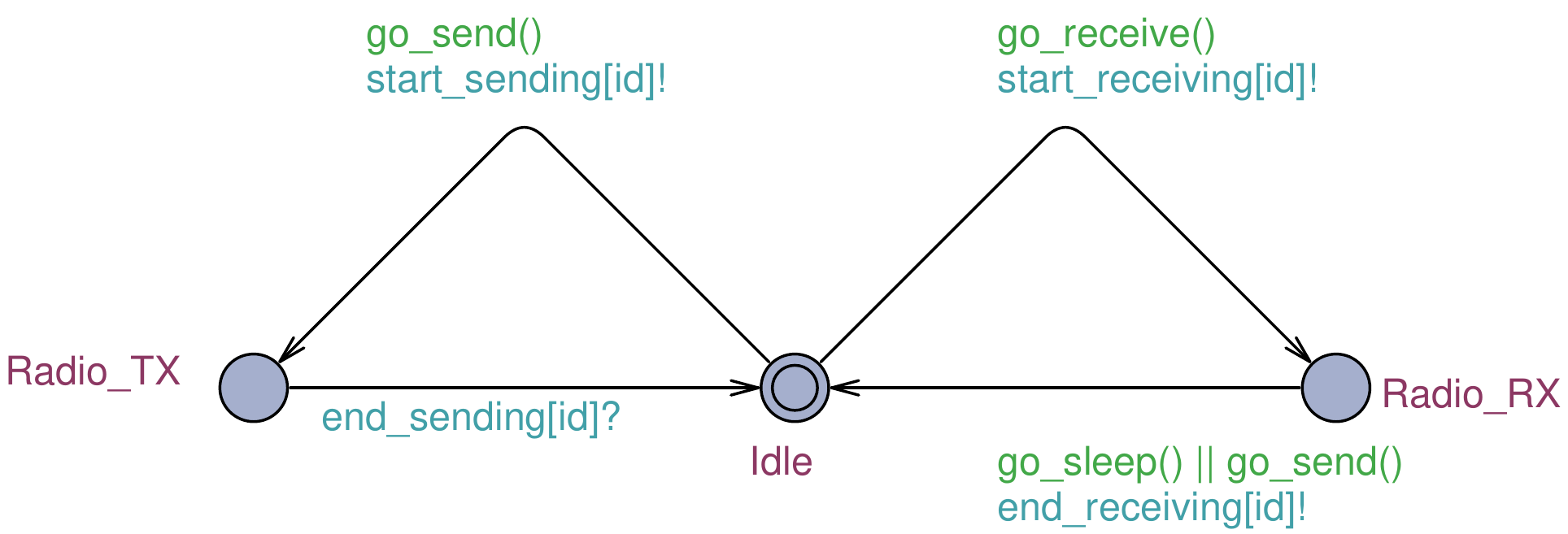}
		\caption{Automaton $\Controller[\id]$}
		\label{fig:wsn-controller}
	\end{center}
	\end{minipage}
	\hspace{0.5cm}
	\begin{minipage}[b]{0.5\linewidth}
{\footnotesize
\begin{verbatim}
bool go_send(){return (r>g)
 ?((csn[id]+1)%C==tsn[id] && clk[id]==k0-(r-g))
 :(csn[id]==tsn[id] && clk[id]==g-r);}

bool go_receive(){return 
(r>0 && 0!=tsn[id] && csn[id]==C-1 && clk[id]==k0-r)
|| (r==0 && 0!=tsn[id] && csn[id]==0)
|| (0<csn[id] && csn[id]<n && csn[id]-1==tsn[id]);}

bool go_sleep(){return csn[id]==n;}
\end{verbatim}
}
\caption{Predicates used in $\Controller[\id]$}
                \label{fig:function defs}
	\end{minipage}
\end{figure}

\paragraph{Controller}
The task of the $\Controller[\id]$ automaton, displayed in Figure~\ref{fig:wsn-controller},
is to put the radio in sending and receiving mode at the appropriate moments.
Figure~\ref{fig:function defs} shows the definition of the predicates used in this automaton.
The radio should be put in sending mode $\Radioswitchtime$ ticks before message transmission starts
(at time $\Guard$ in the transmission slot of $\id$).
If $\Radioswitchtime > \Guard$ then the sender needs to be activated $\Radioswitchtime - \Guard$ ticks before the end of
the slot that precedes the transmission slot.
Otherwise, the sender must be activated at tick $\Guard - \Radioswitchtime$ of the transmission slot.
If the first slot in a frame is an RX slot, then the radio is switched to receiving mode $\Radioswitchtime$ time units
before the start of the frame to ensure that the radio will receive during the full first slot.
However if there is an RX slot after the TX slot then, as described in Section~\ref{sec:radio-switching-time},
the radio is switched to the receiving mode only at the start of the RX slot.
The controller stops the radio receiver whenever either the last active slot has passed or the sender needs to be switched on.

All the channels used in the $\Controller[\id]$ automaton (${\sf start\_sending}$, ${\sf end\_sending}$, ${\sf start\_receiving}$,
${\sf end\_receiving}$ and ${\sf synchronize}$) are urgent, which means that these signals are sent at the moment when the transitions
are enabled.

\begin{figure}[hpt]
	\begin{minipage}[b]{0.5\linewidth}
	\begin{center}
		\includegraphics[width = 9.5cm]{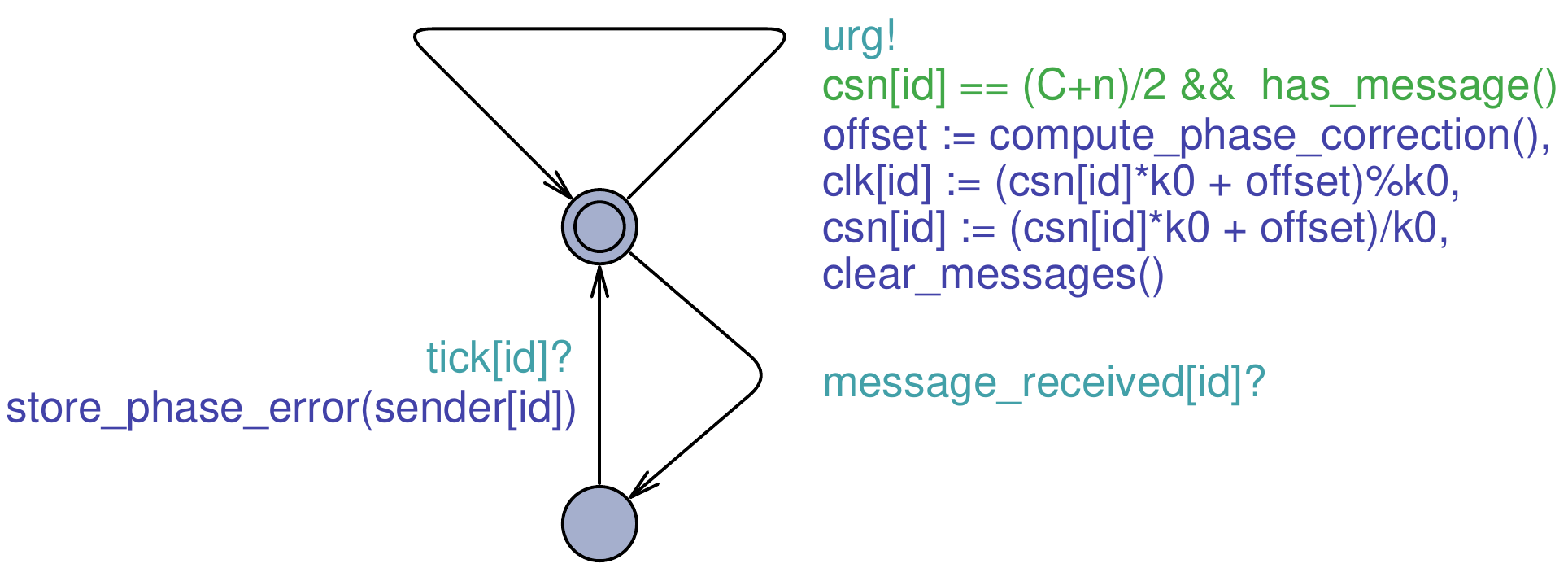}
		\caption{Automaton $\Synchronizer[\id]$}
		\label{fig:wsn-synchronizer}
	\end{center}
	\end{minipage}
	\hspace{0.5cm}
	\begin{minipage}[b]{0.5\linewidth}
{\footnotesize
\begin{verbatim}
        void store_phase_error(int sender)
        {
        phase_errors[msg_counter] =
           (tsn[sender] * k0 + k0 - g) 
                - (csn[id] * k0 + clk[id]);      
        msg_counter++
        }
\end{verbatim}
}
		\caption{Function used in $\Synchronizer[\id]$}
                \label{fig:function defs2}
	\end{minipage}
\end{figure}

\paragraph{Synchronizer} Finally, automaton $\Synchronizer[\id]$ is shown in Figure \ref{fig:wsn-synchronizer}.
The automaton maintains a list of phase differences of all messages received in the current frame, using a local array ${\sf phase\_errors}$.
Local variable ${\sf msg\_counter}$ records the number of messages received.
Whenever the receiver gets a message from a neighboring node (${\sf message\_received}[\id]$), the synchronizer computes and stores
the phase difference using the function ${\sf store\_phase\_error}$ at the next clock tick.
Here the phase difference is defined as the expected time at which
the message transmission ends ({\tt tsn[sender] * k0 + k0 - g}) minus 
the actual time at which the message transmission ends ({\tt csn[id] * k0 + clk[id]}), counting from the start of the frame.
The complete definition is listed in Figure~\ref{fig:function defs2}.
Recall that in our model we abstract from transmission delays.

As explained in Section \ref{sec:synchronization}, the synchronizer computes the value of the phase correction (offset) and adjusts the clock
during the sleeping period of a frame.\footnote{Actually, in the implementation the offset is used to compute the corrected \emph{wakeup time},
that is the moment when the next frame will start \cite{communication}.  In our model we reset the clock, but this should be equivalent.}
Hence, in order to decide in which slot we may perform the synchronization, we need to know the maximal phase difference between two nodes.  
In our model, we assume no joining of networks. When a node receives a message from another node, the phase difference computed using this message will not exceed the length of an active period. Otherwise one of these two nodes will be in sleeping period while the other is sending, hence no message can be received at all. In practice, the number of sleeping slots is much larger than the number of active slots. Therefore it is safe to perform the adjustment in the middle of sleeping period because the desired property described above holds. When the value of {\tt gain} is smaller than $1$  the maximal phase difference will be even smaller.

The function of ${\sf compute\_phase\_correction}$ implements exactly the algorithm listed in Section \ref{sec:synchronization}.

\section{Analysis Results}
\label{analysis-results}
In this section, we present some verification results that we obtained
for simple instances of the model that we described in Section~\ref{sec:model}.
We checked the following invariant properties using the \Uppaal\ model checker:
\begin{verbatim}
INV1 : A[] forall (i: Nodes) forall (j : Nodes) 
   SENDER(i).Sending && neighbor(i,j)imply RECEIVER(j).Receiving

INV2 : A[] forall (i:Nodes) forall (j:Nodes) forall (k:Nodes)
   SENDER(i).Sending && neighbor(i,k) && SENDER(j).Sending && neighbor(j,k)
                   imply i == j

INV3 : A[] not deadlock
\end{verbatim}
The first property states that always when some node is sending, all its neighbors are
listening.  The second property states that never two different neighbors of a given
node are sending simultaneously.  The third property states that the model contains
no deadlock, in the sense that in each reachable state at least one component can make
progress.  The three invariants are basic sanity properties of the gMAC protocol, at least
in a setting with a static topology and no transmission failures.

We used {\Uppaal} on a Sun Fire X4440 machine (with 4 Opteron 8356 2.3 Ghz quad-core processors and 128 Gb DDR2-667 memory)
to verify instances of our model with different number of nodes, different network topologies and different
parameter values.
Table~\ref{tab:resource} lists some of our verification results, including the resources {\Uppaal}
needed to verify if the network is synchronized or not. 
In all experiments, $\C = 10$ and $\K = 29$. 

\begin{table}
	\centering
	\begin{tabular}{|c|c|c|c|c|c|c|c|c|}
		\hline
   		\multirow{2}{*}{$\N/\n$} & \multirow{2}{*}{Topology} &\multirow{2}{*}{$\Guard$} & 
		\multirow{2}{*}{$\Radioswitchtime$} & \multirow{2}{*}{$\frac{\Min}{\Max}$} & 
		\multirow{2}{*}{CPU Time} & \multirow{2}{*} {Peak Memory Usage} & \multirow{2}{*}{Sync}\\
    		&  &  &  &  &  &  & \\
    		\hline
		$3/3$ &  clique & $2$ & $0$ & $1$ & $1.944$ s & $24,180$ KB & YES\\
    		\hline
		$3/3$ &  clique & $2$ & $0$ & $\frac{100,000}{100,001}$ & $492.533$ s & $158,064$ KB & NO\\
    		\hline
		$3/3$ &  clique & $2$ & $1$ & $1$ & $1.976$ s & $68.144$ KB & YES\\
    		\hline
		$3/3$ &  clique & $2$ & $0$ & $\frac{100,000}{100,001}$ & $116.68$ s & $68,144$ KB & NO\\
    		\hline
		$3/3$ &  line & $2$ & $0$ & $1$ & $1.068$ s & $68,144$ KB & YES\\
    		\hline
		$3/3$ &  line & $2$ & $0$ & $\frac{100,000}{100,000}$& $441.308$ s & $68,144$ KB & NO \\
    		\hline
		$3/3$ &  line & $2$ & $1$ & $1$ & $1.041$ s & $68,144$ KB & YES\\
    		\hline
		$3/3$ &  line & $2$ & $1$ & $\frac{100,000}{100,000}$& $99.274$ s & $68,144$ KB & NO \\
    		\hline
    		$3/3$ &  clique & $3$ & $0$ & $1$ & $1.851$ s & $28,040$ KB & YES\\
    		\hline
		$3/3$ &  clique & $3$ & $0$ & $\frac{100,000}{100,001}$ & $575.085$ s & $272,312$ KB & NO\\
    		\hline
		$3/3$ &  clique & $4$ & $0$ & $\frac{350}{351}$ & $115.166$ s & $516,636$ KB & NO\\
    		\hline
		$3/3$ &  clique & $4$ & $0$ & $\frac{351}{352}$ & $147.864$ s & $630,044$ KB & YES\\
    		\hline
    		$3/3$ &  clique & $3$ & $2$ & $1$ & $1.827$ s & $24,184$ KB & YES\\
		\hline
		$3/3$ &  clique & $3$ & $2$ & $\frac{100,000}{100,001}$ & $109.633$ s & $26,056$ KB & NO\\
    		\hline
		$3/3$ &  clique & $4$ & $2$ & $\frac{100,000}{100,001}$ & $533.345$ s & $350,504$ KB & NO\\
    		\hline
		$3/3$ &  clique & $5$ & $2$ & $\frac{587}{588}$ & $72.473$ s & $332,552$ KB & NO\\
    		\hline
		$3/3$ &  clique & $5$ & $2$ & $\frac{588}{589}$ & $99.101$ s & $407,884$ KB & YES\\
    		\hline
		$3/3$ &  clique & $3$ & $5$ & $1$ & $0.076$ s & $21,884$ KB & NO\\
    		\hline
		$3/3$ &  line & $3$ & $0$ & $1$ & $1.05$ s & $23,348$ KB & YES\\
    		\hline
		$3/3$ &  line & $3$ & $0$ & $\frac{451}{452}$ & $29.545$ s & $148,012$ KB & NO\\
    		\hline
		$3/3$ &  line & $3$ & $0$ & $\frac{452}{453}$ & $35.257$ s & $148,012$ KB & YES\\
    		\hline
    		$3/3$ &  line & $3$ & $2$ & $1$ & $1.052$ s & $22,916$ KB & YES\\
		\hline
		$3/3$ &  line & $3$ & $2$ & $\frac{100,000}{100,001}$ & $82.383$ s & $78,360$ KB & NO\\
    		\hline
		$3/3$ &  line & $4$ & $2$ & $\frac{100,000}{100,001}$ & $414.201$ s & $53,752$ KB & NO\\
    		\hline
		$3/3$ &  line & $5$ & $2$ & $\frac{453}{454}$ & $33.16$ s & $147,796$ KB & NO\\
    		\hline
		$3/3$ &  line & $5$ & $2$ & $\frac{454}{455}$ & $38.811$ s & $162,184$ KB & YES\\
    		\hline
		$3/3$ &  line & $3$ & $5$ & $1$ & $0.048$ s & $78,360$ KB & NO\\
    		\hline
		$4/4$ &  clique & $3$ & $0$ & $1$ & $231.297$ s & $1,437,643$ KB & YES\\
    		\hline
		$4/4$ &  clique & $3$ & $0$ & $\frac{450}{451}$ & \multicolumn{3}{|c|}{Memory Exhausted}\\
    		\hline
    		$4/4$ &  clique & $3$ & $2$ & $1$ & $229.469$ s & $1,438,368$ KB & YES\\
		\hline
		$4/4$ &  clique & $3$ & $2$ & $\frac{100,000}{100,001}$ & $14,604.531$ s & $2,317,040$ KB & NO\\
    		\hline
		$4/3$ &  line & $3$ & $0$ & $1$ & $4.749s$ s & $94,748$ KB & YES\\
    		\hline
		$4/3$ &  line & $3$ & $0$ & $\frac{450}{451}$ & \multicolumn{3}{|c|}{Memory Exhausted}\\
    		\hline
		$4/3$ &  line & $3$ & $2$ & $1$ & $4.738$ s & $94,748$ KB & YES\\
    		\hline
		$4/3$ &  line & $3$ & $2$ & $\frac{100,000}{100,001}$ & $1,923.655$ s & $1,264,844$ KB & YES\\
    		\hline
		$5/5$ &  clique & $3$ & $0$ & $1$ & \multicolumn{3}{|c|}{Memory Exhausted}\\
    		\hline
		$5/5$ &  clique & $3$ & $2$ & $1$ & \multicolumn{3}{|c|}{Memory Exhausted}\\
    		\hline
		$5/3$ &  line & $3$ & $0$ & $1$ & $46.54$ s & $249,976$ KB & YES\\
    		\hline
		$5/3$ &  line & $3$ & $2$ & $1$ & $46.489$ s & $250,880$ KB & YES\\
    		\hline
		$6/3$ &  line & $3$ & $0$ & $1$ & $508.19$ s & $2,316,416$ KB & YES\\
    		\hline
		$6/3$ &  line & $3$ & $2$ & $1$ & $502.871$ s & $2,317,040$ KB & YES\\
    		\hline
		$7/3$ &  line & $3$ & $0$ & $1$ & \multicolumn{3}{|c|}{Memory Exhausted}\\
    		\hline
		$7/3$ &  line & $3$ & $2$ & $1$ & \multicolumn{3}{|c|}{Memory Exhausted}\\
    		\hline	
	\end{tabular}
	\caption{Model checking experiments}
	\label{tab:resource}
\end{table}

Clearly, the values of network parameters, in particular clock parameters $\Min$ and $\Max$, affect the result of the verification.
Table~\ref{tab:resource} shows several instances where the protocol is correct for perfect clocks ($\Min = \Max$) but fails
when we decrease the ratio $\frac{\Min}{\Max}$.
It is easy to see that the protocol will always fail when $\Radioswitchtime\ge\Guard$.
Consider any node $i$ that is not the last one to transmit within a frame.
Right after its sending slot, node $i$ needs $\Radioswitchtime$ ticks to get its radio into receiving mode.
This means that --- even with perfect clocks --- after $\Guard$ ticks another node already has started sending
even though the radio of node $i$ is not yet receiving. 
Even when $\Radioswitchtime < \Guard$, the radio switching time has a clear impact on correctness: the
larger the radio switching time is, the larger the guard time has to be in order to ensure correctness.
Using {\Uppaal}, we can fully analyze line topologies with at most seven nodes if all clocks are perfect.
For larger networks \Uppaal\ runs out of memory. 
A full parametric analysis of this protocol will be challenging, also due to the
impact of the network topology and the selected slot allocation.
Using \Uppaal, we discovered that for certain topologies and slot allocations the Median algorithm may
always violate the above correctness assertions, irrespective of the choice of the guard time.
For example, in a $4$ node-network with clique topology and $\Min$ and $\Max$ of $100.000$ and $100.001$, respectively,
if the median of the clock drifts of a node becomes $-1$, the median algorithm divides it by $2$ and generates
$0$ for clock correction value and indeed no synchronization happens.
If this scenario repeats in three consecutive time frames for the same node, that node runs $\Guard =3$ clock cycles behind and gets out of sync.

\begin{figure}[hpt]
   \begin{center}
                \includegraphics[width = 15cm]{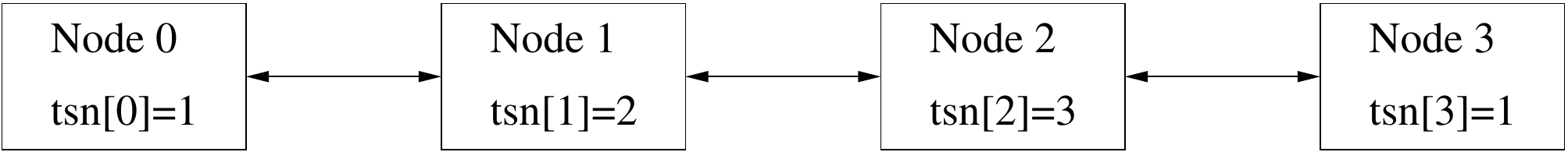}
                \caption{A problematic network configuration}
                \label{fig:counterexample}
        \end{center}
\end{figure}

Another example in which the algorithm may fail is displayed in Figure~\ref{fig:counterexample}.
This network has 4 nodes, connected by a line topology, that send in slots 1, 2, 3, 1, respectively.
Since all nodes have at most two neighbors, the Median algorithm prescribes that nodes will
correct their clocks
based on the first phase error that they see in each frame.  For the specific topology and slot
allocation of Figure~\ref{fig:counterexample}, this means that
node 0 adjusts its clock based on phase errors of messages it gets from node 1,
node 1 adjusts its clock based on messages from node 0,
node 2 adjusts its clock based on messages from node 3, and
node 3 adjusts its clock based on messages from node 2.
Hence, for the purpose of clock synchronization, we have two disconnected networks!
Thus, if the clock rates of nodes 0 and 1 are lower than the clock rates of
nodes 2 and 3 by just an arbitrary small margin, then two subnetworks will eventually get out of sync.
These observations are consistent with results that we obtained using  \Uppaal.
If, for instance, we set $\Min[\id] = 99$ and $\Max[\id] = 100$, for all nodes $\id$ then neither {\tt INV1} nor {\tt INV2} holds.
In practice, it is unlikely that the above scenario will occur due to the fact that in the implementation slot allocation
is random and dynamic.  Due to regular changes of the slot allocation, with high probability node 1 and node 2 will now and then
adjusts their clocks based on messages they receive from each other.

\begin{figure}[hpt]
   \begin{center}
                \includegraphics[width = 12cm]{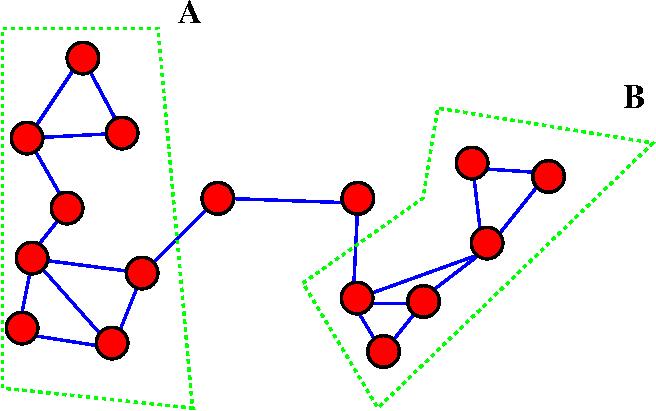}
                \caption{Another problematic network configuration with two communities}
                \label{fig:counterexample2}
        \end{center}
\end{figure}

However, variations of the above scenario may occur in practice, even in a setting with dynamic slot allocation.
In fact, the above synchronization problem is also not restricted to line topologies.
We call a subset $C$ of nodes in a network a \emph{community} if each node in $C$ has more neighbors within $C$ than outside $C$ \cite{Newman04}.
For \emph{any} network in which two disjoint communities can be identified,
the Median algorithm allows for scenarios in which these two parts become unsynchronized.
Due to the median voting mechanism, the phase errors of nodes outside a community will not affect the nodes within this community,
independent of the slot allocation.
Therefore, if nodes in one community $A$ run slow and nodes in another community $B$ run fast then
the network will become unsynchronized eventually, even in a setting with infinitesimal clock drifts.
Figure~\ref{fig:counterexample2} gives an example of a network with two communities.
\begin{figure}[h!]
	\centering
	\includegraphics[scale=0.8]{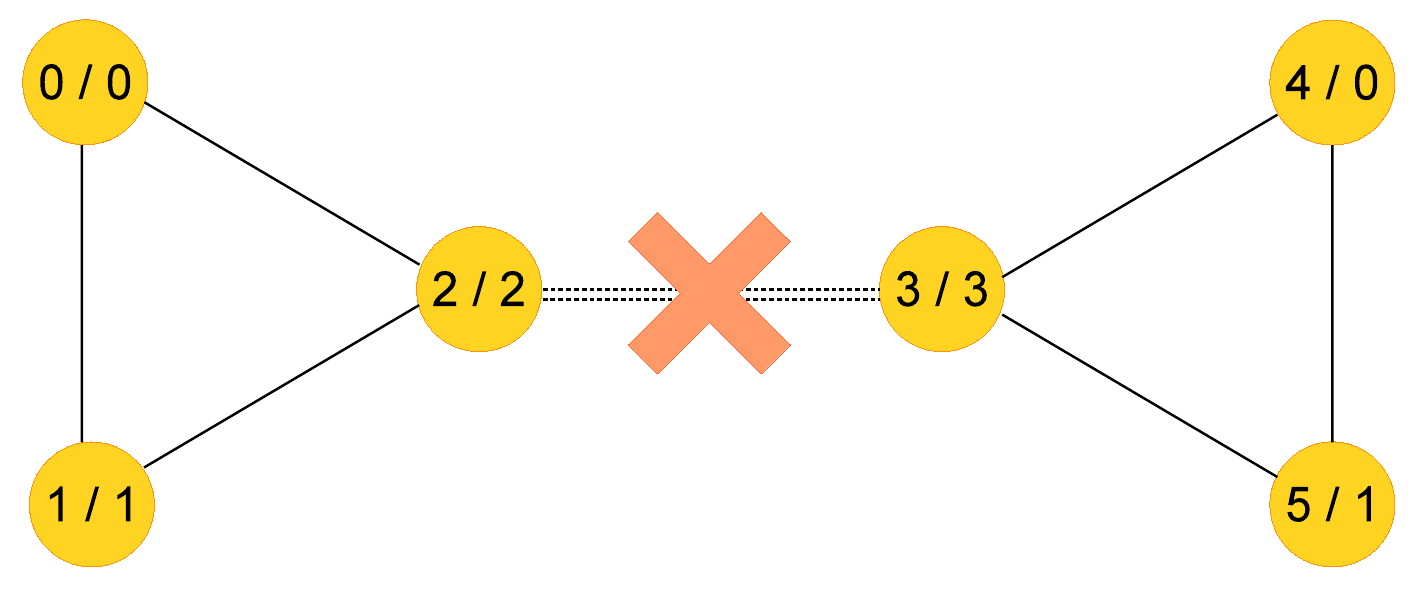} 
	\caption{A network with two communities that we analyzed using \Uppaal}
	\label{fig:2communities}
\end{figure}

Using \Uppaal, we succeeded to analyze instances of the simple network with two communities displayed in Figure \ref{fig:2communities}.
The numbers on the vertices are the node identifiers and the transmission slot numbers, respectively.
Table~\ref{tab:2communities} summarizes the results of our model checking experiments.
\begin{table}
	\centering
	\begin{tabular}{|c|c|c|c|c|c|}
		\hline
   		\multirow{2}{*}{$\Guard$} & \multirow{2}{*}{$\Radioswitchtime$} & 
		Fast Clock & Slow Clock &
		\multirow{2}{*}{CPU Time} & \multirow{2}{*} {Peak Memory Usage}\\
    		&  & Cycle Length & Cycle Length &  & \\
		\hline
		$2$ & $0$ & $1$& $1$ & \multicolumn{2}{|c|}{Memory Exhausted}\\
		\hline
		$2$ & $0$ & $99$& $100$ & $457.917$ s & $2,404,956$ KB \\
		\hline
		$2$ & $1$ & $99$& $100$ & $445.148$ s & $2,418,032$ KB \\
		\hline
		$3$ & $0$ & $99$& $100$ & $416.796$ s & $2,302,548$ KB \\
		\hline
		$3$ & $2$ & $1$& $1$ & \multicolumn{2}{|c|}{Memory Exhausted} \\
		\hline 
		$3$ & $2$ & $99$& $100$ & $22.105$ s & $83,476$ KB \\
		\hline
		$3$ & $2$ & $451$& $452$ & $798.121$ s & $3,859,104$ KB \\
		\hline
		$3$ & $2$ & $452$& $453$ & \multicolumn{2}{|c|}{Memory Exhausted} \\
		\hline 
		$4$ & $0$ & $99$& $100$ & $424.935$ s & $2,323,004$ KB \\
		\hline
		$4$ & $1$ & $99$& $100$ & $464.503$ s & $2,462,176$ KB \\
		\hline
		$4$ & $2$ & $99$& $100$ & $420.742$ s & $2,323,952$ KB \\
		\hline	
	\end{tabular}
	\caption{Model checking experiments of a network with two communities}
	\label{tab:2communities}
\end{table}

We still need to explore how realistic our counterexamples are.  We believe that network topologies with multiple communities
occur in many WSN applications.  Nevertheless, in practice the gMAC protocol appears to perform quite well for static
networks.  It might be that problems do not occur so often in practice due to the probabilistic distributions of
clock drift and jitter.

\section{Conclusions}
\label{conclusions}

We presented a detailled \Uppaal\ model of relevant parts of the clock synchronization algorithm that is currently being used
in a wireless sensor network that has been developed by Chess \cite{del52,communication}.
The final model that we presented here may look simple, but the road towards this
model was long and we passed through numerous intermediate versions on the way.
Using \Uppaal,  we established that in certain cases a static, fully synchronized network may eventually become unsynchronized if
the current Median algorithm is used, even in a setting with infinitesimal clock drifts.

In \cite{HSV09}, we proposed a slight variation of the gMAC algorithm that does not have the correctness problems of the
Median algorithm.  However, our algorithm still has to be tested in practice.
Assegei \cite{Assegei08} proposed and simulated three alternative algorithms, to be used instead of the  Median algorithm,
in order to achieve decentralized, stable and energy-efficient synchronization of the Chess gMAC protocol.
It should be easy to construct \Uppaal\ models for Assegei's algorithms: basically, we only have to modify the definition
of the {\tt compute\_phase\_correction} function.
Recently, Pussente \& Barbosa \cite{Pussente09}, also proposed a very interesting new clock synchronization algorithm
--- in a somewhat different setting --- that achieves an $O(1)$ worst-case skew between the logical clocks of neighbors.
Much additional research is required to analyze correctness and performance of these algorithms in the realistic
settings of Chess with large networks, message loss, and network topologies that change dynamically.
Starting from our current \Uppaal\ model, it should be relatively easy to construct models
for the alternative synchronization algorithms in order to explore their properties.

Analysis of clock synchronization algorithms for wireless sensor networks is an extremely challenging area
for quantitative formal methods.  One challenge is to come up with the right abstractions that will
allow us to verify larger instances of our model.  Another challenge is to make more detailled (probabilistic)
models of radio communication and to apply probabilistic model checkers and specification tools
such as PRISM \cite{PRISM2} and CaVi \cite{FehnkerFM09}.

Several other recent papers report on the application of \Uppaal\  for the analysis of protocols for
wireless sensor networks, see e.g.\ \cite{FehnkerHM07,FehnkerFM09,TschirnerXY08,HSV09}.
In \cite{WoehrleLT09}, \Uppaal\ is also used to automatically test the power consumption of wireless sensor networks.
Our paper confirms the conclusions of \cite{FehnkerHM07,TschirnerXY08}: despite the small number of nodes that
can be analyzed, model checking provides valuable insight in the behavior of protocols for wireless sensor networks,
insight that is complementary to what can be learned through the application of simulation and testing.

\paragraph{Acknowledgement}
We are most grateful to Frits van der Wateren for his patient explanations of the subtleties of the gMAC protocol.
We thank Hern\'{a}n  Bar\'{o}  Graf for spotting a mistake in an earlier version of our model, and Mark Timmer
for pointing us to the literature on communities in networks.
Finally, we thank the anonymous reviewers for their comments.

\bibliographystyle{eptcs}
\bibliography{abbreviations,dbase,reference}
\end{document}